\documentclass{pasj01}

\Received{$\langle$reception date$\rangle$}
\Accepted{$\langle$acception date$\rangle$}
\Published{$\langle$publication date$\rangle$}

\RequirePackage{color}

\newcommand{\lsim}{\hspace{0.3em}\raisebox{0.4ex}{$<$}\hspace{-0.75em}\raisebox{-.7ex}{$\sim$}\hspace{0.3em}}

\begin{document}

\title{The Milky Way Tomography with Subaru Hyper Suprime-Cam. I. Halo substructures}

\author{
Yoshihisa~Suzuki\altaffilmark{1},
Masashi~Chiba\altaffilmark{1}, 
Yutaka~Komiyama\altaffilmark{2}, 
Kohei~Hayashi\altaffilmark{3,1}, 
Masayuki~Tanaka\altaffilmark{4,5}, 
Tetsuya~Fukushima\altaffilmark{4},
Scott~G.~Carlsten\altaffilmark{6},
Akira~Tokiwa\altaffilmark{7},
Tian~Qiu\altaffilmark{7}
and Masahiro~Takada\altaffilmark{7}
}

\altaffiltext{1}{Astronomical Institute, Tohoku University, Aoba-ku,
Sendai 980-8578, Japan \\E-mail: {\it yoshihisa.suzuki@astr.tohoku.ac.jp}}
\altaffiltext{2}{Department of Advanced Sciences, Faculty of Science and Engineering,
Hosei University, 184-8584 Tokyo, Japan}
\altaffiltext{3}{National Institute of Technology, Sendai College, 48 Nodayama, Medeshima-Shiote, Natori-shi, Miyagi 981-1239, Japan}
\altaffiltext{4}{National Astronomical Observatory of Japan, 2-21-1 Osawa, Mitaka,
Tokyo 181-8588, Japan}
\altaffiltext{5}{The Graduate University for Advanced Studies, Osawa 2-21-1, Mitaka,
Tokyo 181-8588, Japan}
\altaffiltext{6}{Department of Astrophysical Sciences, Princeton University, Peyton Hall, Princeton NJ 08544 USA}
\altaffiltext{7}{Kavli Institute for the Physics and Mathematics of the Universe, Kashiwa, Chiba, 277-8583, Japan}


\maketitle

\begin{abstract}
We analyze the photometric data in the Wide layer of the Hyper Suprime-Cam Subaru Strategic Program (HSC-SSP) over $\sim 1,200$ deg$^{2}$ to uncover new halo substructures beyond the distance, $D_{\odot}\sim$ 30 kpc, from the Sun. For this purpose, we develop an isochrone filter for an old, metal-poor stellar system to extract the faint main-sequence stars at a range of distances. With this method, we detect, not only the previously discovered substructures such as the Orphan Stream, but also the new overdensity toward Bo\"otes at about $D_{\odot}\sim$ 60 kpc and the new stream-like feature toward Pisces at around $D_{\odot}\sim$ 60 kpc. It has been suggested that a small-scale overdensity exists in this direction of Pisces (the so-called Pisces Overdensity), but our results show that the overdensity is widely spread with a tidally elongated feature. Combining our results with the ongoing Hyper Suprime-Cam narrow-band survey and the near-future spectroscopic survey with Prime Focus Spectrograph (PFS) will allow us to place strong constraints on the origin of these halo substructures.
\end{abstract}
\section{Introduction}

Stars in the Milky Way (MW) halo hold much fundamental information on how a galaxy like our own formed. These stars, identified first with high radial velocities \citep{Oort1922, Oort1926} and later with weak metal lines \citep{Roman1950} near the Sun, show a different chemo-dynamical structure from those in the luminous disk component, where most of the stars reside. The quest for the origin of these halo stars has opened up Galactic Archaeology studies for constructing a possible scenario of Galaxy formation based on their kinematical and chemical properties since the seminal paper by \citet{Eggen1962}. While these authors proposed a picture of a monolithic, free-fall dissipative collapse of a proto-Galactic cloud, the advent of newly analyzed datasets in the halo population has revealed a more complex picture of developing the MW halo through accretion and merging of smaller stellar systems \citep{Searle1978}, which actually accords with the modern picture of galaxy formation. 
 

A currently standard scenario for galaxy formation is based on Cold Dark Matter models in the $\Lambda$-dominated Universe ($\Lambda$CDM). In this context, galaxies are thought to have been formed by the repeated merging and accretion of the smaller objects through gravitational interaction (e.g., \citet{White1978}). The numerical simulations based on the $\Lambda$CDM model have shown that the relics of the past formation history should still be inscribed in the halo \citep{Bekki2001, Bullock2005, Cooper2010}. Several lines of evidence for these past accretion events are actually observed in the halo of nearby galaxies in the form of halo streams and substructures (e.g., \citet{Martinez2010, Ferguson2016}). These studies have provided significant insight into our understanding of the galaxy formation process.

For the MW, thanks to the advent of wide-field systematic photometric and spectroscopic surveys such as the Sloan Digital Sky Survey (SDSS) and the precise astrometry survey with {\it Gaia}, our understanding of the MW structure, especially in the inner part of the MW halo at the heliocentric distance of $D_{\odot} \lsim 30$ kpc, has been advanced significantly (e.g., \citet{York2000, Gaia2016}). These modern data allow us to carry out the reconstruction of the past merging history of halo building blocks based on high-precision 6D phase-space information and chemical abundance patterns of halo stars (e.g., \citet{Helmi2018, Belokurov2018}).

On the other hand, our understanding of the MW outer halo beyond $D_{\odot}\sim$ 30 kpc is still far from complete. This is mainly because the halo substructures to be found there are characterized as low surface brightness, thereby being difficult to detect \citep{Shipp2023, Horta2023}. Although many works have been performed using intrinsically bright stars such as the horizontal branch stars and RR Lyrae variables as a halo tracer, it is not straightforward to find halo substructures because of the small number of available sample stars. In fact, in order to search for halo structures with sufficient statistical significance, the wide and deep photometric survey of abundant faint stars along the main-sequence is desired. 
	
Here, we make use of the photometric data obtained from the Subaru Strategic Program using the Hyper Suprime-Cam (HSC), hereafter called HSC-SSP. By virtue of its wide footprint ($\sim$ 1,200 $\mathrm{deg^2}$) and deep photometry ($\lesssim$ 26.5 mag in the $i$-band with 5$\sigma$ depth), it is possible to search for new substructures at the outer part of the MW halo. In this paper, we report on the first detection of a candidate stellar stream at around $D_{\odot} \sim$ 60 kpc toward Pisces and also the global overdensity toward Bo\"otes at that distance. In Section 2, we show the photometric data for our analysis and then describe the methodology for the detection of distant halo substructures. In Section 3, we show our results and discuss the origin of the candidate halo substructure in Section 4. Finally, our conclusion is drawn in Section 5. In this paper, magnitudes are given in the AB system \citep{Oke1983}.

\section{Data and Method}
We first briefly introduce the survey design of the HSC-SSP, which is the basis for the current work. We next show how to extract candidate stellar objects as many as possible by using the color index information, then explain how to select halo stars to use the following analysis. Finally, we introduce an isochrone filter to extract many of main-sequence stars located at a certain range of distances.

\subsection{HSC-SSP}
The HSC-SSP is a large imaging survey that began in March 2014 and was completed in January 2022, a period of about 7.5 years. The observation has been conducted using the HSC \citep{Miyazaki2018} installed on the prime focus of the 8.2 m Subaru Telescope. Over the course of $\sim$ 330 nights, this program has acquired photometric data with various optical filters and depths. Here, we utilize the photometric data observed in the Wide layer, composed of three fields: Spring, Fall, and North (see Table \ref{table1}). In total, it covers $\sim 1200$ deg$^2$ in five photometric bands ($g$, $r$, $i$, $z$, and $y$) with $5\sigma$ limiting magnitude ($26.5^{+0.2}_{-0.2}$, $26.5^{+0.2}_{-0.2}$, $26.2^{+0.2}_{-0.3}$, $25.2^{+0.2}_{-0.3}$ and $24.4^{+0.2}_{-0.3}$) \citep{Aihara2022}. The HSC-SSP catalog (S21A) was generated by the hscPipe 8.4, updated with an FGCM code (hscPipe 8.5.3), based on the pipeline developed for Vera C. Rubin Observatory \citep{Ivezic2008, Juric2017, Bosch2019} and calibrated against Pan-STARRS1 photometry and astrometry \citep{Schlafly2012, Tonry2012, Magnier2013, Chambers2016}.

\begin{table*}[t]
\caption{The HSC-SSP survey footprint in the Wide Layer. The observed areas in the equatorial and Galactic coordinates for the Spring, North, and Fall fields are summarized. Over the course of 7.5 years, $\sim$ 1200 $\rm{deg}^{2}$ has been swept into the sky in this program.}

\begin{centering}
\begin{tabular}{lccccc}
\hline
Field &   RA  &  DEC  &  l  &  b  & area \\
      & (deg) & (deg) & (deg) & (deg) &  (deg$^2$)  \\
\hline \hline
Spring&  [127.5, 225.0]                    & [-2.0, 5.0]  & [225.0, 360.0] & [20.0, 65.0] & $\sim$ 680 \\
North &  [200.0, 250.0]                    & [42.0, 44.5] & [70.0, 90.0] & [40.0, 75.0] & $\sim$ 125 \\
Fall  &  [330.0, 360,0] \& [0.0, 30.0]     & [-1.0, 7.0]  & [60.0, 150.0] & [-65.0,-35.0] & $\sim$ 395 \\
\hline
Wide layer & &       &        &      &  $\sim$ 1200 \\
\hline

\end{tabular}
\label{table1}

\end{centering}
\end{table*}

\begin{figure*}[h!]
\begin{center}
\includegraphics[width=160mm]{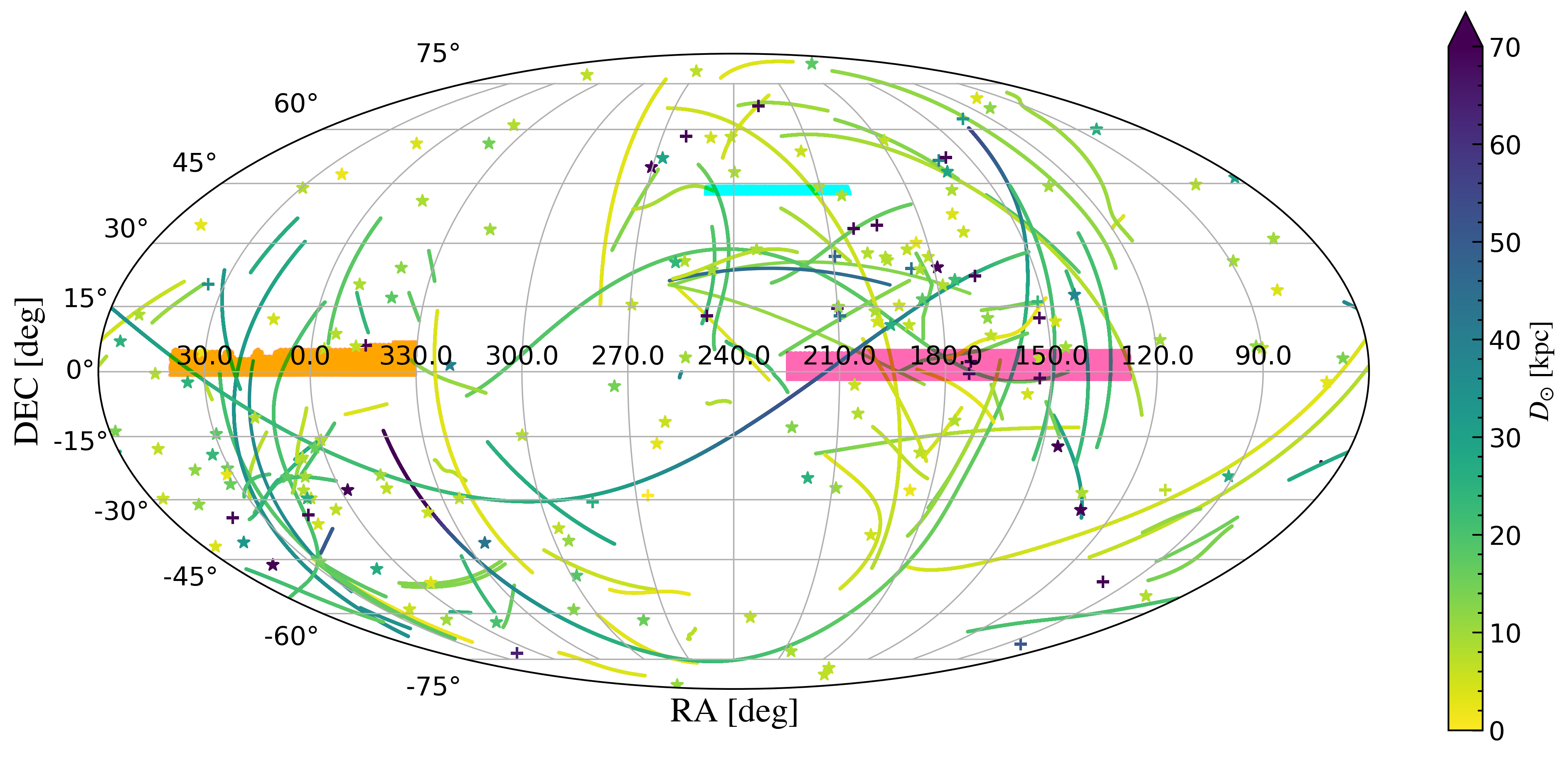}
\end{center}
\hspace*{10mm}
\caption{
The HSC-SSP footprint in the Wide layer in the equatorial coordinate. The pink, orange, and cyan-colored areas correspond to the Spring, Fall, and North fields, respectively. The symbols show previously discovered dwarf galaxies (plus) and globular clusters (star). The continuous lines are the stellar streams already found. The color represents the distance from the Sun, being close (yellow) to far (purple), sequentially. From this figure, while we can see that many stellar streams have been identified within 30 kpc of the Sun, the deeper HSC data is expected to allow us to examine farther out than that.
}

\label{Fig1}
\end{figure*}

In Fig.\ref{Fig1}, we show the three fields in the Wide layer in the equatorial coordinate system (J=2000.0). Areas painted in pink, orange, and cyan are Spring, Fall, and North fields, respectively. In addition, we overplot the distribution of previously discovered dwarf galaxies \citep{McConnachie2012}, globular clusters \citep{Harris1996}, and stellar streams  \citep{Mateu2023} in the MW, represented as symbols of plus, stars, and continuous lines, respectively. The colors range from yellow to purple, meaning increasing distance from the Sun. Most of the known stellar streams are seen within $D_{\odot}\sim$ 30 kpc from the Sun, whereas beyond $D_{\odot}\sim$ 30 kpc, only the dwarf galaxies are detected. However, numerical experiments have demonstrated that more coherent stellar streams originated from low- and intermediate-mass stellar progenitors can be found beyond $D_{\odot}\sim$ 30 kpc \citep{Horta2023}. Since the photometric data obtained with HSC-SSP can capture faint main-sequence stars within about 100 kpc, it is highly expected that stellar streams at such large distances will be detected with this dataset.

\subsection{Selection of halo stars}
First, we extract objects that are considered to be point sources; in the HSC-SSP catalog, each source is binary classified as a point source or extended source by the ${\it i\_extendedness\_value}$: point source (as value 0) and extended source (as value 1). This criterion is determined by whether the $i$-band photometric magnitude difference is larger (point-source) or smaller (extended-source) than -0.018 for the cases between using the point spread function (PSF) model and using the model that combines the PSF model and typical galaxy surface brightness such as an exponential profile in photometry. Note that the HSC-SSP uses the best-seeing ($\sim$ 0.6 arcsec) $i$-band images as the first priority for source detection, which is especially sensitive for the weak-lensing survey.

\begin{figure}[h!]
\includegraphics[width=80mm]{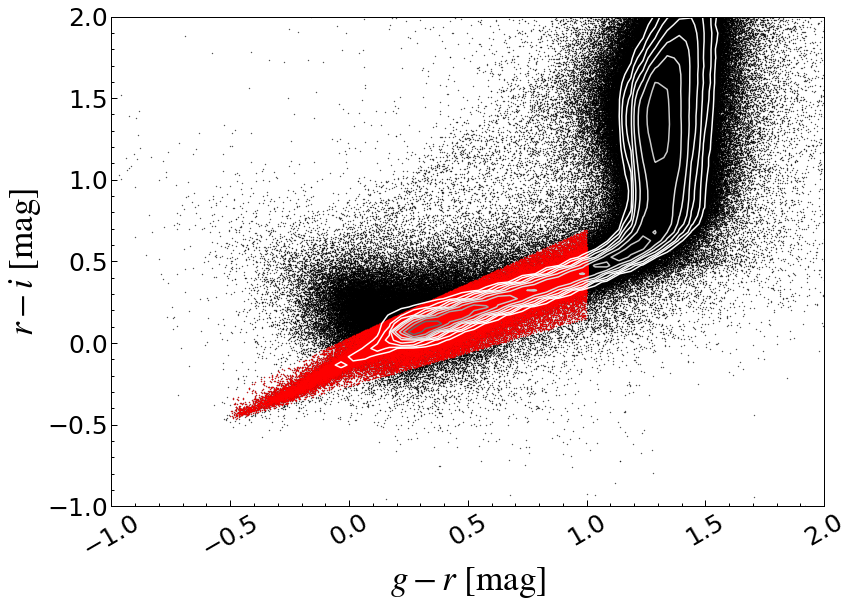} 
\hspace*{10mm}
\caption{
The distribution of objects considered as point sources on the color-color diagram ($g$-$r$, $r$-$i$) for those brighter than 22 mag in the $i$-band. We plot them as black dots. The gray-scale contours show the number distribution on a common logarithmic scale from $10^{3}$ to $10^{5}$, corresponding to the stellar locus. In the following analysis, the point sources within the region indicated by red dots are considered halo stars.
}

\label{Fig2}
\end{figure}

We then extract as many stars as possible from the obtained data. To do so, we sample objects brighter than 24.5 mag in the $i$-band photometry based on the PSF model. This is because the probability that the selected point sources are correctly distinguished as stars or galaxies is less than 50\% for objects fainter than $i$=24.5. This is based on the result of cross-matching the objects considered as point sources by the HSC-SSP catalog with those after distinguishing stars and galaxies by Hubble observations obtained in the COSMOS region of the HSC-SSP footprint \citep{Aihara2018}.

After the major contaminants are removed, we make a further selection using the color-color diagram. In Fig.\ref{Fig2}, we show the distribution of objects identified as ‘pure’ point sources on the color-color diagram ($g$-$r$, $r$-$i$). For clarification, we plot the objects brighter than 22 mag in $i$-band. Besides we overplot the gray-scale contours from $10^{3}$ to $10^{5}$ in number on the common logarithmic scale. In this space, we can first identify a sequence where the objects are most concentrated, called the stellar locus. On this sequence, there is a bend around ($g$-$r$, $r$-$i$) $\sim$ (1.3, 0.5), which is known to be mainly due to the presence of many M- and K-type stars in the thin disk. There is also a component near ($g$-$r$, $r$-$i$) $\sim$ (0.0, 0.2), which deviates from the stellar locus. This component is composed of quasars and star-forming galaxies, as revealed by SDSS spectroscopic follow-up observations \citep{Ahumada2020, Lyke2020}.

Based on these properties of point sources in this diagram, we extract the most likely halo stars based on the following three aspects. First, we remove objects with $g-r>$ 1.0 to remove M- and K-type stars in the thin disk. Next, we remove quasars and galaxies with active star formation outside the main stellar locus. Finally, we extract only the sources that are inside the stellar locus. As a result, the sources brighter than $i$=24.5 within the region indicated by the red dots in Fig.\ref{Fig2} are regarded as halo stars in the following analysis.

\subsection{Isochrone-filter method}
 In order to effectively find substructures in the halo of the MW, we make an isochrone filter based on a theoretical stellar evolution model \citep{Bressan2012}. The results of this publicly available calculation, however, do not exactly cover the case for the HSC photometry. Therefore we use an empirical linear equation for converting SDSS-based filters to HSC-based filters for the spectroscopic follow-up stars using one-color index information as deduced in \cite{Akiyama2018} as follows, under the assumption that this relation is applicable to all the sky:

\begin{center}
\begin{equation}
  \left(	
  \begin{array}{c}
    g_{\mathrm{HSC}} \\
    r_{\mathrm{HSC}} \\
    i_{\mathrm{HSC}} \\
    z_{\mathrm{HSC}} \\
    y_{\mathrm{HSC}}
  \end{array}
  \right)
  = A
  \left(
  \begin{array}{c}
    g_{\mathrm{SDSS}} \\
    r_{\mathrm{SDSS}} \\
    i_{\mathrm{SDSS}} \\
    z_{\mathrm{SDSS}} \\
    y_{\mathrm{SDSS}}
  \end{array}
  \right)
  +
  \left(
  \begin{array}{c}
    -0.011 \\
    -0.001 \\
     0.003 \\
    -0.006 \\
     0.003
   \end{array}
  \right)
\end{equation}
\end{center}
\begin{center}
\begin{equation}
A = 
\left(
  \begin{array}{ccccc}
    0.926 & 0.074 & 0 & 0 & 0 \\
    0 & 0.996 & 0.004 & 0 & 0 \\
    0 & -0.016 & 1.106 & 0 & 0 \\
    0 & 0 & 0.006 & 0.094 & 0 \\
    0 & 0 & -0.419 & 1.419 & 0
   \end{array}
\right)
\end{equation}
\end{center}

We then determine the width of the filter to cover the stellar population. Under the plausible assumption that most of the stellar systems in the outer halo are metal-poor and old, we adopt the theoretical isochrone with an age of 13.5 Gyr and metallicity [M/H] of -1.8. Also, to reduce the influence of foreground bright stars, our isochrone filter is focused and limited to the extraction of faint main-sequence stars. Then, the width of the filter is given at each 0.2 mag interval between $i$=18 and 24 mag, taken from the median photometric error in the corresponding $i$-band magnitude. The metallicity distribution is expressed with a shift of about 0.1 magnitude in the color index ($g$-$i$) at $i$=24.5 mag. Thus, our isochrone filter and its width take into account the dispersion of the intrinsic metallicity (-2.1$<$[M/H]$<$-1.3) for old main-sequence stars with $\gtrsim$ 10 Gyr. 

In order to apply the filter to the HSC-SSP photometry, we first consider the dust extinction using the maps calculated in \cite{Schlegel1998}. Then we make isochrone filters with distance moduli from 17.1 mag ($D_{\odot}\sim 26.3$ kpc) to 19.3 ($D_{\odot}\sim 72.4$ kpc) mag as evenly shifted from top to bottom on the color-magnitude diagram. As for the distance error associated with the finite width of the filter, it ranges from 10\% ($D_{\odot} \sim$ 25 $\mathrm{kpc}$) to 28\% ($D_{\odot} \sim$ 60 $\mathrm{kpc}$).


\begin{figure*}[h!]
\begin{center}
\includegraphics[width=160mm]{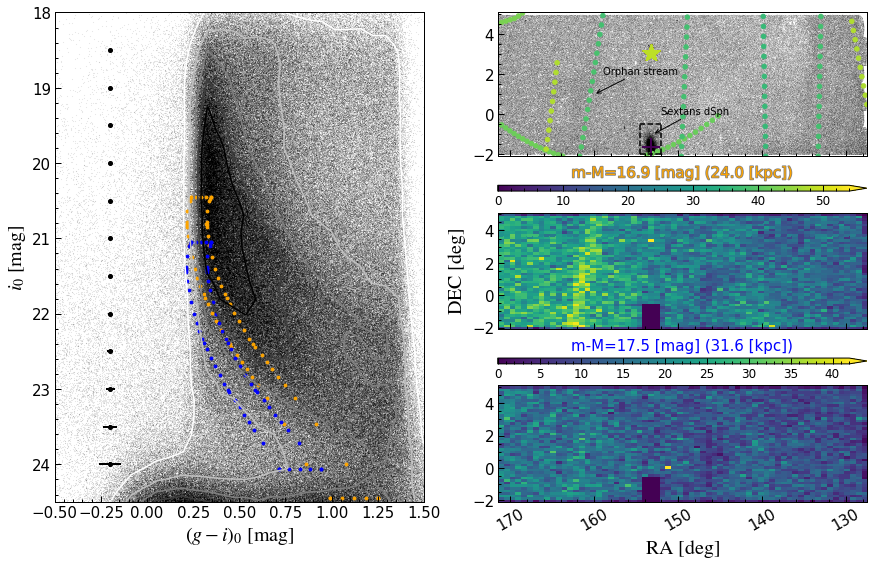}
\end{center}
\hspace*{10mm}
\caption{
The application of the isochrone-filter method to the Spring field. (left) The halo stars (black dots) in the color-magnitude diagram $((g-i)_{0}, i_{0})$. The typical photometric error in each $i$-band bin is shown as black circles with error bars. (right) The top panel shows the spatial distribution of all the halo stars (black dots) in this field. We also add the previously discovered dwarf galaxies, globular clusters, and stellar streams as the same symbols in Fig.\ref{Fig1}. The middle and bottom panels show the number of stars within each pixel ($\Delta$RA=0.6875 deg, $\Delta$DEC=0.16 deg) after isochrone-filtering in the left panel shown with orange and blue dotted regions, respectively. The corresponding distance modulus and distance are ((m-M)=16.9 mag, $D_{\odot}\sim$ 24.0 kpc) and (m-M)=17.5 mag, $D_{\odot}\sim$ 31.6 kpc), respectively. For the star count, we remove the Sextans dSph surrounded in the dashed region in the top panel. From the middle panel, we can clearly see the Orphan stream over (RA, DEC) = ($162^\circ$, -$2^\circ$) to ($160^\circ$, $4^\circ$), indicating that this technique is valid for stellar stream detection.
}
\label{Fig3}
\end{figure*}

\section{Results}
In each subsection for the Spring, North, and Fall fields below, we summarize the general characteristics of the field and explain the distribution of stars in the color-magnitude diagram. We especially focus on only the notable feature in the spatial distribution of stars inside an isochrone filter at a selected range of distances from the Sun, whereas in Appendix 1, we show all the spatial distributions of stars in each field while varying distance moduli with a small interval.

\begin{figure*}[h!]
\begin{center}
\includegraphics[width=160mm]{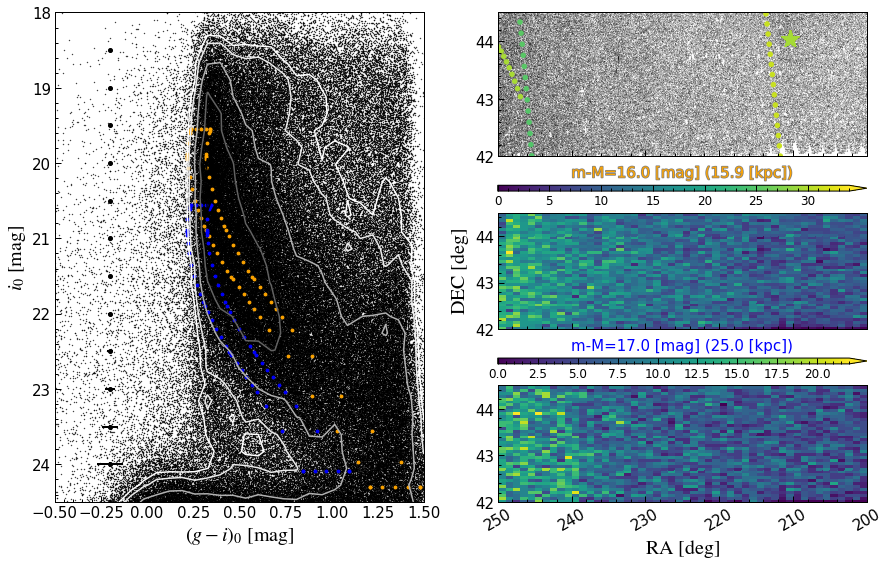}
\end{center}
\hspace*{10mm}
\caption{
The application of the isochrone-filter method to the North field in the same manner as shown in Fig.\ref{Fig3}. The orange and blue dotted isochrones in the left panel pick out the stars at ((m-M)=16.0 mag, $D_{\odot}\sim$ 15.9 kpc) and ((m-M)=17.0 mag, $D_{\odot}\sim$ 25.0 kpc), respectively, and the right middle and bottom panels show the corresponding distribution of stars within each pixel ($\Delta$RA=0.6875 deg, $\Delta$DEC=0.16 deg). As shown in the text, the color-magnitude diagram exhibits the presence of a previously undiscovered structure at $i$ = 23$\sim$24 mag, though is hardly spatially dense. This suggests the presence of a faint overdense structure in this direction, for which we dub "Bo\"otes Overdensity".
}
\label{Fig4}
\end{figure*}

\subsection{Spring field}
 The Spring field is the most crowded area with substructures, because of the presence of the Sagittarius stream at $D_{\odot} \sim$ 40 kpc: this stream reveals many main sequence stars in the color-magnitude diagram at corresponding distance modulus, thereby making it difficult to investigate other substructures. Therefore, we perform a substructure search for the region excluding such regions.

In Fig.\ref{Fig3}, we summarize the characteristic features of stars in the Spring field. In the left panel, all the halo stars in the color-magnitude diagram ($(g-i)_0$, $i_0$) are shown as black points. Note that the subscript '0' with filters means that dust extinction correction has been performed. Also, the median photometric error for each object is indicated by a black circle with an error bar at each step of 0.5 mag between $i$=18.5 and 24 mag.

 Focusing on the main-sequence stars in this panel, we remark that the foreground and the outer halo stars can be distinguished owing to the small photometric error. Therefore, it is sufficient to apply an isochrone filter only to the abundant main-sequence stars in the color-magnitude diagram compared to the use of bright stars such as RR Lyrae, blue horizontal-branch (BHB), and red giant branch (RGB) stars.
 
In the right panels of Fig.\ref{Fig3}, we show the spatial distribution of halo stars in the Spring field. The top panel shows the spatial distribution of all the halo stars in this field. The middle and bottom panels show the number density in each pixel ($\Delta$RA=0.6875 deg, $\Delta$DEC=0.16 deg) after isochrone-filtering with a distance modulus of (m-M)=16.9 mag ($D_{\odot}\sim$ 24.0 kpc) and 17.5 mag ($D_{\odot}\sim$ 31.6 kpc), respectively, corresponding to stars in the orange and blue dashed regions in the left panel. We note that we do not normalize the star counts, so the range of the color bar is different in each panel. For the number counts, we mask the known dwarf galaxy, Sextans dSph at (RA, DEC) = ($162^\circ$, $-2^\circ$). As is clear, the previously discovered stellar stream, the Orphan stream, is confirmed over (RA, DEC) = ($162^\circ$, $-2^\circ$) to ($160^\circ$, $4^\circ$) at a distance of about $D_\odot = 24$ kpc (right middle panel). This demonstrates that the method using our isochrone filter is effective for exploring substructures in the outer halo. However, except this known stream, there are no notable substructures, where the density variation of stars is confined only within 1 sigma over the field\footnote{Some local overdensities are present around Sextans dSph, including a globular cluster and dwarf candidate as reported elsewhere.}.

\begin{figure*}[h!]
\begin{center}
\includegraphics[width=160mm]{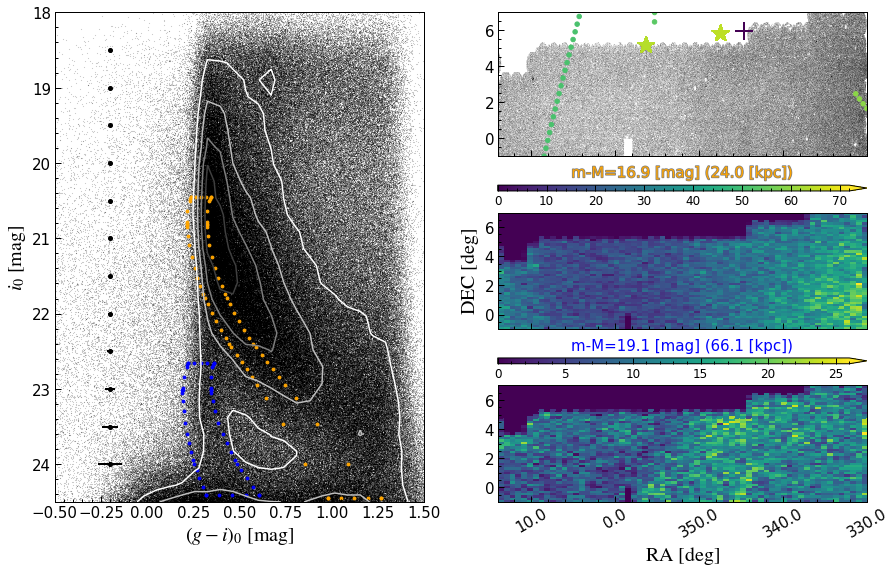}
\end{center}
\hspace*{10mm}
\caption{
The application of the isochrone-filter method to the Fall field in the same manner as shown in Fig.\ref{Fig3}. The orange and blue dotted isochrones in the left panel pick out the stars at ((m-M)=16.9 mag, $D_{\odot}\sim$ 24.0 kpc) and ((m-M)=19.1 mag, $D_{\odot}\sim$ 66.1 kpc), respectively, and the right middle and bottom panels show the corresponding distribution of stars within each pixel ($\Delta$RA=0.6875 deg, $\Delta$DEC=0.16 deg). From both the color-magnitude diagram and the spatial distribution of the stars, we can see the stream-like feature toward Pisces over 330 $<$ RA $<$ 360 at around 60 kpc from the Sun  (right bottom panel).
}
\label{Fig5}
\end{figure*}

\subsection{North field}

The North field has only a few known stellar streams that are located approximately within a few 10 kpc from the Sun based on the combination of the stellar population and kinematics by Gaia satellite \citep{Ibata2019}. Since these nearby streams do not affect our search for streams in the outer halo beyond $D_{\odot}\sim$ 30 kpc, we apply the isochrone filter without any masking.

We show the result of this application to the North field in Fig.\ref{Fig4}. In the left panel, we plot the color-magnitude diagram for all the halo stars and the typical photometric errors in the same way as in Fig.\ref{Fig3}. In the right panels, we show the distribution of the halo stars in the North field. The top panel shows the spatial distribution of all the halo stars. The middle and bottom panels show the number density within each pixel of ($\Delta$RA=0.6875 deg, $\Delta$DEC=0.16 deg) after applying the isochrone filters with the distance modulus of (m-M)= 16.0 mag ($D_{\odot}\sim$ 15.9 kpc) and 17.0 mag ($D_{\odot}\sim$ 25 kpc), respectively, corresponding to stars in the orange and blue dashed regions in the left panel. 

Compared to the Spring field, there exist no apparent stream-like substructures, except for the following two features. We recognize from the color-magnitude diagram that the signature of a rather large-scale substructure appears at distances larger than 30 kpc from the Sun. This structure, however, is not clearly seen as a high-density stellar stream but in the form of widely spread overdensity over 240 $<$ RA $<$ 250 deg in the spatial distribution (right middle and bottom panels of Fig.4). From this characteristic, we call it as “the Bo\"otes overdensity”, which appears to be present over $D_{\odot}\sim$ 60 kpc (see Appendix 1 for the sequential distribution of stars at different distant moduli). In this series of our papers, we attempt to investigate the possible origin of this overdensity from the viewpoints of the kinematics and chemical abundance of the K-type, RR Lyrae, or BHB stars at the same distance (Suzuki et al., in preparation). 

As seen in the right middle and bottom panels of Fig.\ref{Fig4}, there appears yet another overdensity at around RA $\sim$ 210 deg. Although this overdensity is rather weak in these panels, this feature is seen in all the isochrone filters with different distance moduli as shown in Appendix 1 (See Figure \ref{Fig9}), whereas it disappears when the selection of stars is limited to bright ones of i$<$24 mag (see Figure \ref{Fig11}). Therefore, we doubt that this overdensity is due to contamination such as galaxies. To confirm whether this overdensity is true or not, in the near future, it will be possible to utilize the data obtained by the Euclid Satellite, which will be able to carry out star-galaxy separation more precisely taking advantage of observations from space (e.g., \citet{Laureijs2011}). Since this North region overlaps with the planned observation area of Euclid, the identification of this overdensity will be eagerly performed.

\subsection{Fall field}

The Fall field, like the Spring field, is a region where the many main-sequence stars from the Sagittarius stream are dominant at around $D_{\odot}\sim$ 30 kpc. Also, a dwarf galaxy such as IC 1613 is present. Therefore, we apply the isochrone filter by excluding these regions.

We show the result of this application to the Fall field in Fig.\ref{Fig5}. In the left panel, we plot the color-magnitude diagram for all the halo stars and the typical photometric errors in the same manner as in Fig.\ref{Fig3}. In the right panels, we show the distribution of halo stars in the Fall field. The top panel shows the spatial distribution of all the halo stars. The middle and bottom panels show the number density within each pixel ($\Delta$RA=0.6875 deg, $\Delta$DEC=0.16 deg) after applying the isochrone filters with the distance modulus of (m-M)=16.9 mag ($D_{\odot} \sim$ 24.0 kpc) and 19.1 mag ($D_{\odot} \sim$ 66.1 kpc), respectively, corresponding to stars in the orange and blue dashed regions in the left panel. For the star count in this field, we set to mask the region bounded by RA=($358^\circ$, $359^\circ$) and DEC=($-1.0^\circ$, $0.0^\circ$). This is because the analysis using only stars near the main-sequence turn-off has suggested the local number density excess, indicating a new dwarf galaxy candidate. We summarize its structural parameters and physical quantities in this series of papers (Suzuki et al., in preparation). 

As seen in the right middle and bottom panels, we can identify the highly overdense structure over 330 $<$ RA $<$ 360 deg, or more specifically extended from (RA, DEC)=($345^\circ \sim 360^\circ$, $0^\circ$) to (RA, DEC)=($330^\circ \sim 345.6^\circ$, $5^\circ$). This may be a part of the so-called Herclues-Aquila Cloud (HAC) \citep{Belokurov2007} at the distance below 30 kpc (right middle panel). A recent orbital analysis of RR Lyrae stars obtained with the Gaia DR2 dataset has shown that the apogalactic distances of the HAC member stars are in the range of 15-25 kpc, i.e., do not exceed $D_{\odot}$ $\sim$ 50 kpc \citep{Simion2018}. However, our analysis indicates that this structure is extended up to the distance of 60 kpc from the Sun, where the influence of the HAC is negligible (right bottom panel), thereby suggesting that these stream-like features are discovered for the first time thanks to the deep HSC photometry. The density contrast at this distance scale is apparently high: while the number density of stars per pixel ($\Delta$RA=0.6875 deg, $\Delta$DEC=0.16 deg) is almost 0 at around RA$\sim360^\circ$, this overdensity shows the significantly larger number density of stars of around 30. The existence of this substructure is also apparent in the color-magnitude diagram in the left panel, and also in Appendix 1 (Figure \ref{Fig10}) beyond $D_{\odot}\sim$ 30 kpc. We can identify the concentration of the main-sequence stars, and also confirm the signature of the RGB stars, which appears to continuously connect with the main-sequence stars distributed along a single isochrone.

\begin{figure}[h!]
\begin{center}
\includegraphics[width=85mm]{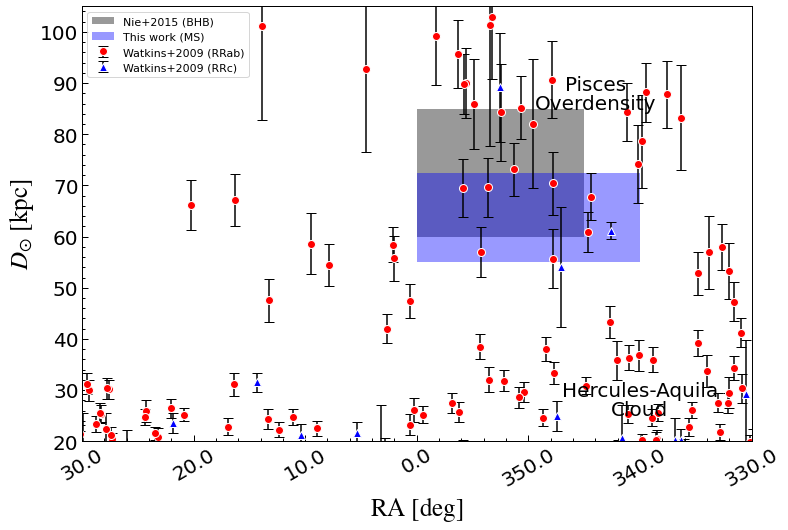}
\end{center}
\hspace*{10mm}
\caption{
The schematic view of the substructures in the Fall field. The depth structure of the Pisces Overdensity, which is clarified by the faint main-sequence stars in this study, is shown by the blue-filled area; the distribution of RR Lyrae stars, where the existence of the Pisces Overdensity has first been noted, is shown by red dots (RRab) and blue triangles (RRc), respectively. The presence of Helcules-Aquila Cloud around 30 kpc from the Sun and that of the Pisces Overdensity above 50 kpc from the Sun are indicated by the clustering of RR Lyrae stars. Also, the estimated spatial distribution of the Pisces Overdensity using blue-horizontal branch stars is shown by the gray-filled area. This figure shows that the distance from the Sun of the Pisces overdensity derived in this study is almost consistent with the previous studies, while its projected spatial distribution is more extended than before.
}
\label{Fig6}
\end{figure}

\begin{figure*}[h!]
\begin{center}
\includegraphics[width=160mm]{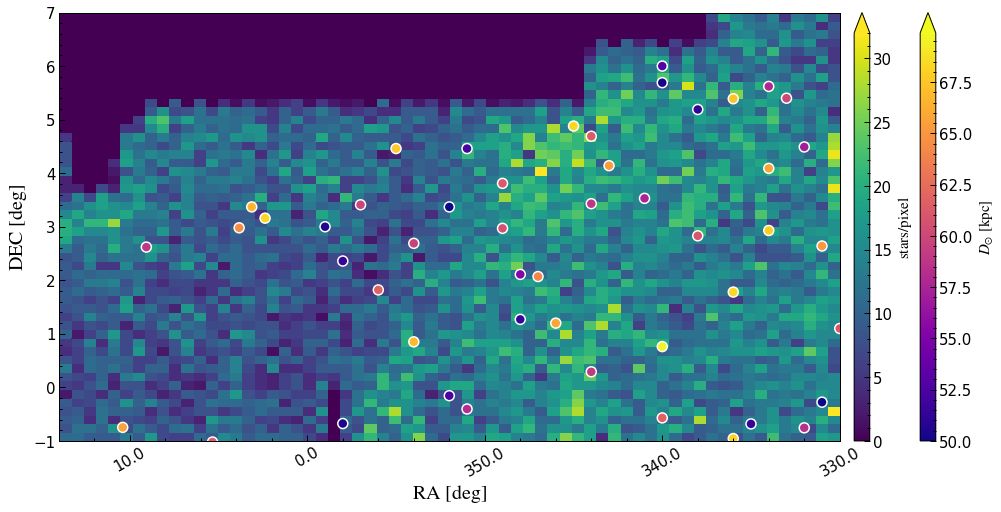}
\end{center}
\hspace*{10mm}
\caption{
The stellar stream candidate at the distance modulus of 18.9 mag ($D_{\odot} \sim$ 66 kpc) and the distribution of the independently selected Blue Stragglers (BSs) in the Fall region using the HSC catalog. The circles in the figure indicate BSs with sequential colors corresponding to the distance from the Sun. This figure shows that there is a correspondence between the stream-like feature and the BS distribution.
}
\label{Fig7}
\end{figure*}

\section{Discussion}
We first discuss the property of the morphologically stream-like candidate identified in the Fall field and its possible origin. The prospect for probing large-scale overdensity appeared in the North field, named "Bo\"otes Overdensity" here, is also discussed. 

The feature reported here in the Fall field can be related to the previously discovered substructure: Pisces Overdensity. It is pointed out by \cite{Watkins2009} that the distribution of RR Lyrae stars is somewhat concentrated between $D_{\odot}\sim$ 60 and 100 kpc (see Figure \ref{Fig6}), and the metallicity [Fe/H] estimated as about -1.5. Based on the direction of the overdensity, they named it Pisces Ovedensity. However, it was not clear what is the origin of the concentration of these few tens of RR Lyrae stars. Since then, several observational analyses have been conducted, resulting in the following, three scenarios to explain the origin of the Pisces Ovedensity.

The first scenario is proposed by \cite{Nie2015}. They conducted the SCUSS survey, where using the $u$-band filters, they selected the blue horizontal-branch (BHB) stars as the standard candle. From the number distribution of BHB stars (Figure \ref{Fig6}), they argued that the progenitor of the Pisces Ovedensity is a small dwarf galaxy like Carina dwarf. However, there is no direct evidence to support this progenitor galaxy.

The second scenario is suggested by \cite{Deason2018}. They use the public data obtained by Subaru/HSC. By using the $z$-band filters, they effectively extracted the BHB stars as the distance indicator. From the distribution of the BHB stars, they first discovered ‘the Pisces Plume’, which was named after the stretched feature from the Sun. Besides they conclude that the Pisces Overdensity originated from the Large Magellanic Cloud (LMC) from the orbital analysis.
 
 The third scenario is advocated by \cite{Chandra2023a} based on the H3 survey \citep{Conroy2019}. By using multiple photometry, they selected the RGB stars as the distance indicator. From the spectroscopic follow-up by using the Magellan Inamori Kyocera Echelle (MIKE), they first mentioned that the progenitor is the same as that producing the Gaia Enceladus Sausage (hereafter GE/S, \citep{Helmi2018, Belokurov2018}) from the orbital analysis in the H3 footprint. A recent work by \cite{Chandra2023b} shows the relationship among these three scenarios: The origin of the Pisces Overdensity is tidal debris of the GE/S, whereas the Pisces Plume originates from tidal debris of the LMC. They argued that there is no correlation between them. 

Our result is complementary to these previous studies. That is, instead of using only a few tens of the BHB, RR Lyrae, and RGB stars, our work is based on the abundant main-sequence stars at a certain range of distances like what SDSS photometry has done (e.g., \citet{Belokurov2006}). From our result, it is suggested that the Pisces Overdensity is much wider than expected (Figure \ref{Fig6}). Therefore, there is a possibility that the Pisces Plume and Pisces Ovedensity are related to each other. To constrain the nature of the progenitor more tightly, we need wider photometric and spectroscopic surveys because of the spatial spread of the Pisces Overdenisty. 

We corroborate the existence of this extended overdensity that we have identified in this region by comparing with the distribution of Blue Straggler stars (BSs), which have been selected based on HSC ${\it griz}$ photometry to select BHBs with the same A-type spectra, as reported elsewhere (Fukushima et al. prep., see also \cite{Fukushima2019} for the selection method of BSs and BHBs from the HSC photometric data). The result is shown in Fig.\ref{Fig7}, for which the distribution of halo stars is derived from the isochrone filter with the distance modulus 18.9 mag ($D_{\odot} \sim$ 66 kpc). The color code for the number density of stars is the same as used in Fig.\ref{Fig3}. The circles show the most likely BS candidates, for which the probability is higher than 0.8 based on the extended Bayesian analysis.  Also, we limit the BS sample to the distance between $D_{\odot}$=50 kpc and 70 kpc to match the distribution of the discovered stream-like feature located around 60 kpc. As is clear from Fig.\ref{Fig7}, the distribution of these independently selected BSs and that of the main-sequence stars are well overlapped with each other. This indicates that the progenitor of this overdensity is a metal-poor, old stellar system.

To get further insight into this overdensity in the context of the formation of the MW, we have the following prospects using the Subaru Telescope. First, the photometric mapping of this overdensity is planned by using the narrow band filter $NB395$ for HSC, which is designed to cover Ca II H\&K absorption lines as a measure of stellar metallicities\footnote{This observation is a part of our program for hunting extremely metal-poor stars with the use of NB395, called ZERO (Zero Enrichment Rare Objects) survey.}. By combining $NB395$ with the $g$ and $i$ band photometry from the HSC-SSP, we can estimate the [Fe/H] for each star. 

Second, a direct spectroscopic follow-up of the stars in the overdensity will be important to obtain not only their metallicities but also their chemical abundance patterns. This will be possible with Subaru Prime Focus Spectrograph (PFS), which will allow us a wide-field and massively multiplexed spectroscopic capability. Therefore, we will obtain a more detailed picture of the origin of the Pisces Overdenisty.  

In summary, the substructure discovered in the Fall field is almost overlapped with Pisces Overdensity, but its stream-like feature is first discovered in this study. Moreover, this candidate substructure is extended than that found by \cite{Nie2015}. It is likely that Pisces Overdensity is a part of a large stellar stream having a more extended structure. Also, as shown in Fig.7, this structure associates the enhancement of the BSs: BSs are thought to be abundant in old stellar systems \citep{Piotto2004}. Therefore, we conclude that this candidate halo substructure is considered to be a relatively old stellar system.

Regarding the North field, there have been no previous reports on the overdensity toward B\"ootes at around the distance of 60 kpc from the Sun. To further constrain the origin of this substructure, we plan to conduct a spectroscopic follow-up observation for the K-giant stars extrapolated from the isochrone on the color-magnitude diagram. Then the combination with the proper motions from the Gaia DR3 will enable us to perform a chemo-dynamical analysis for the halo overdensity. In particular, it will be possible to investigate whether this halo substructure is associated with the tidally disruption of the GE/S event or not.


\section{Conclusion}
In order to detect new halo substructures in the outer halo regions of the MW, we have analyzed the photometric data of the HSC-SSP Wide layer covering over $\sim$ 1,200 $\mathrm{deg}^2$ in the sky. For this purpose, we have focused on the abundant, faint main-sequence stars available from HSC, which allow us to identify and analyze distant halo structures with sufficient statistical significance, in contrast to the studies using sparse RR Lyr or BHB stars. Also, motivated by the fact that many of the known stellar streams originate from globular clusters or dwarf spheroidal galaxies, we have developed the isochrone filter for such an old, metal-deficient stellar system to extract the halo structure consisting of many main-sequence stars at a range of large distances from the Sun.

With this method, we have found a part of the previously discovered substructure, Pisces Overdensity at around $D_{\odot}\sim$ 60 $\mathrm{kpc}$ from the Sun. Wide and deep HSC data in this region show, for the first time, that this substructure reveals a clear stream-like feature, suggesting that it is associated with the tidal disruption of a stellar system. Furthermore, this halo substructure is overlapped with the spatial distribution of the BSs, i.e., the sample of old, metal-poor stars, which therefore suggests such a stellar population for the progenitor stellar system. In addition to this substructure toward Pisces, we have also identified the global overdensity toward Bo\"otes, which should exist at around 60 $\mathrm{kpc}$.
Further constraints on these halo substructures, based on the acquisition of photometric or spectroscopic metallicity as well as kinematics, will enable us to get insight into the origin of its progenitor galaxy.

The HSC-SSP Wider layer that we have focused on halo substructures in this paper also includes more information on the MW halo, such as the global density profile in the various directions and yet unidentified star clusters and satellites, which will be addressed in our next papers.

\begin{ack}
This work is supported in part by JSPS Grant-in-Aid for Scientific Research (B) (No. 25287062) and MEXT Grant-in-Aid for Scientific Research (No. 16H01086, 17H01101 and 18H04334 for MC, No. 18H04359 and No. 18J00277 for KH). TM is supported by Grant-in-Aid for JSPS Fellows (No. 18J11326). The HSC collaboration includes the astronomical communities of Japan and Taiwan and Princeton University.  The HSC instrumentation and software were developed by the National Astronomical Observatory of Japan (NAOJ), the Kavli Institute for the Physics and Mathematics of the Universe (Kavli IPMU), the University of Tokyo, the High Energy Accelerator Research Organization (KEK), the Academia Sinica Institute for Astronomy and Astrophysics in Taiwan (ASIAA), and Princeton University.  Funding was contributed by the FIRST program from the Japanese Cabinet Office, the MEXT, JSPS,  JST,  the Toray Science  Foundation, NAOJ, Kavli IPMU, KEK, ASIAA, and Princeton University. This paper makes use of software developed for the Large Synoptic Survey Telescope. We thank the LSST Project for making their code freely available. The Pan-STARRS1 (PS1) Surveys have been made possible through contributions of the Institute for Astronomy, the University of Hawaii, the Pan-STARRS Project Office, the Max-Planck Society and its participating institutes, the Max Planck Institute for Astronomy and the Max Planck Institute for Extraterrestrial Physics, The Johns Hopkins University, Durham University, the University of Edinburgh, Queen's University Belfast, the Harvard-Smithsonian Center for Astrophysics, the Las Cumbres Observatory Global Telescope Network Incorporated, the National Central University of Taiwan, the Space Telescope Science Institute, the National Aeronautics and Space Administration under Grant No. NNX08AR22G issued through the Planetary Science Division of the NASA Science Mission Directorate, the National Science Foundation under Grant No.AST-1238877, the University of Maryland, and Eotvos Lorand University (ELTE).
\end{ack}



\appendix

\section{A tomographic view of the HSC-SSP footprint}
\begin{figure*}[h!]
\begin{center}
\includegraphics[width=150mm]{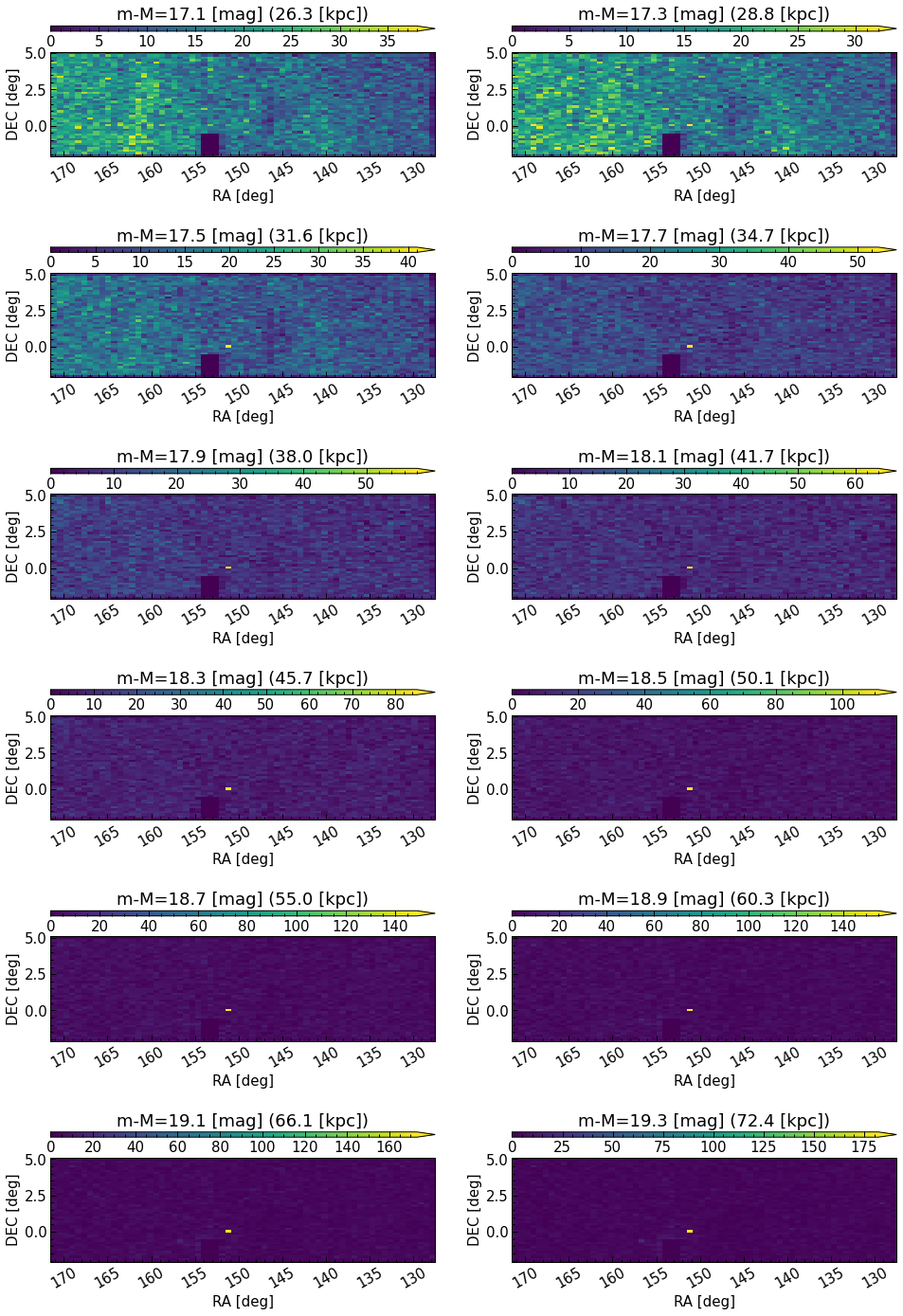}
\end{center}
\hspace*{10mm}
\caption{
Stellar number density distribution in the Spring field after our isochrone-filter selection. The filter is the same as in Fig.\ref{Fig3}. The distance modulus is varied from 17.1 (26.3 kpc) to 19.3 (72.4 kpc) in 0.2 mag increments to be equally spaced on the color-magnitude diagram. The color bar indicates the number density, meaning that the number contained in each pixel increases from purple to yellow. Note that the color bar here shows absolute numbers that are not normalized throughout. This figure shows that in the Spring field, the Orphan stream is dominant within 30 kpc of the sun, while beyond that point, the distribution is smooth except around Sextans.
}
\label{Fig8}
\end{figure*}

\begin{figure*}[h!]
\begin{center}
\includegraphics[width=150mm]{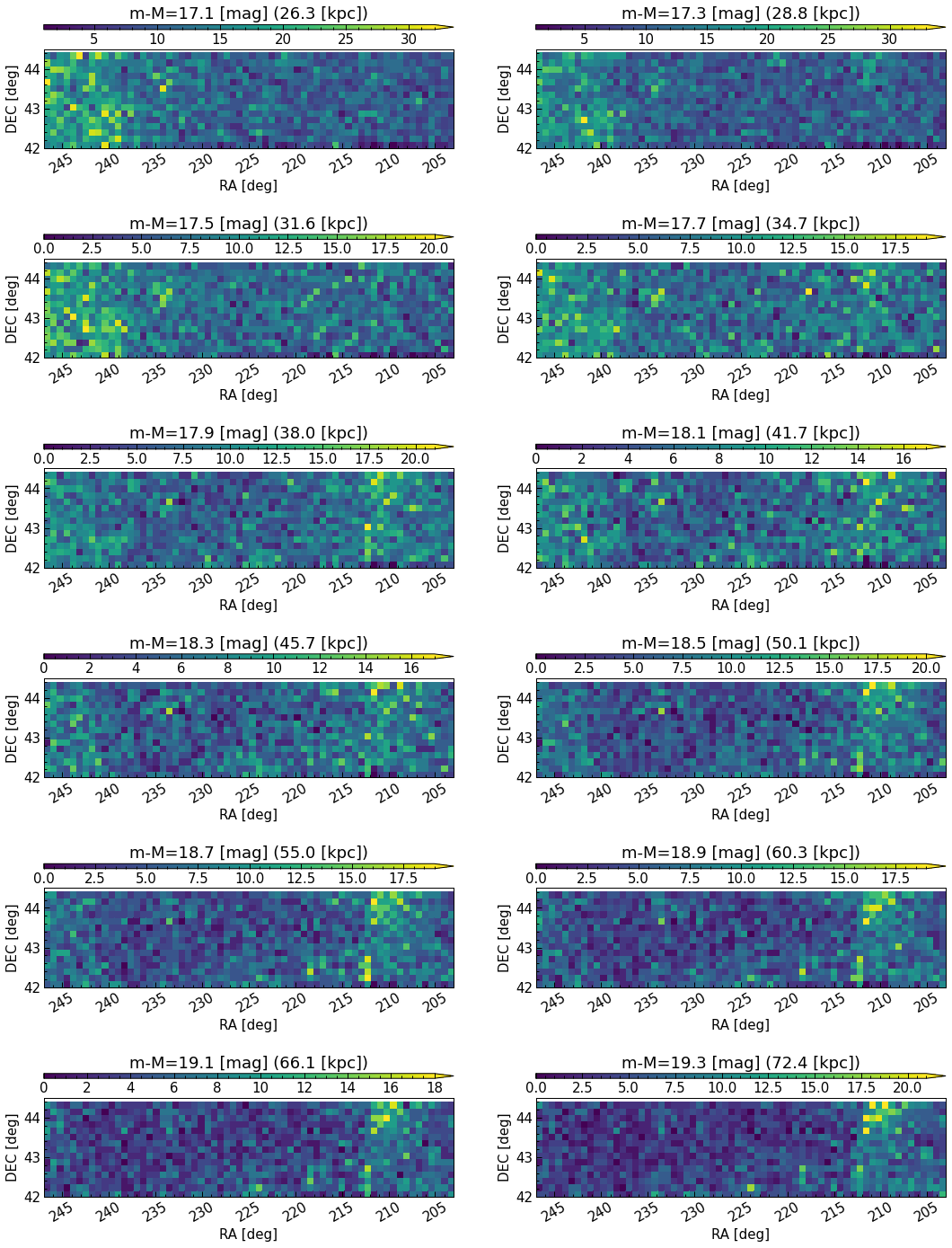}
\end{center}
\hspace*{10mm}
\caption{
Stellar number density distribution in the North field after our Isochrone-filter selection. The panels are shown in the same way as Fig.\ref{Fig8}. This figure shows that in the North field, there are no stream-like features. Note that there is an overdensity at around RA $\sim$ 210 deg, but this might be due to contamination such as galaxies as clearly shown in Appendix 2.
}
\label{Fig9}
\end{figure*}

\begin{figure*}[h!]
\begin{center}
\includegraphics[width=150mm]{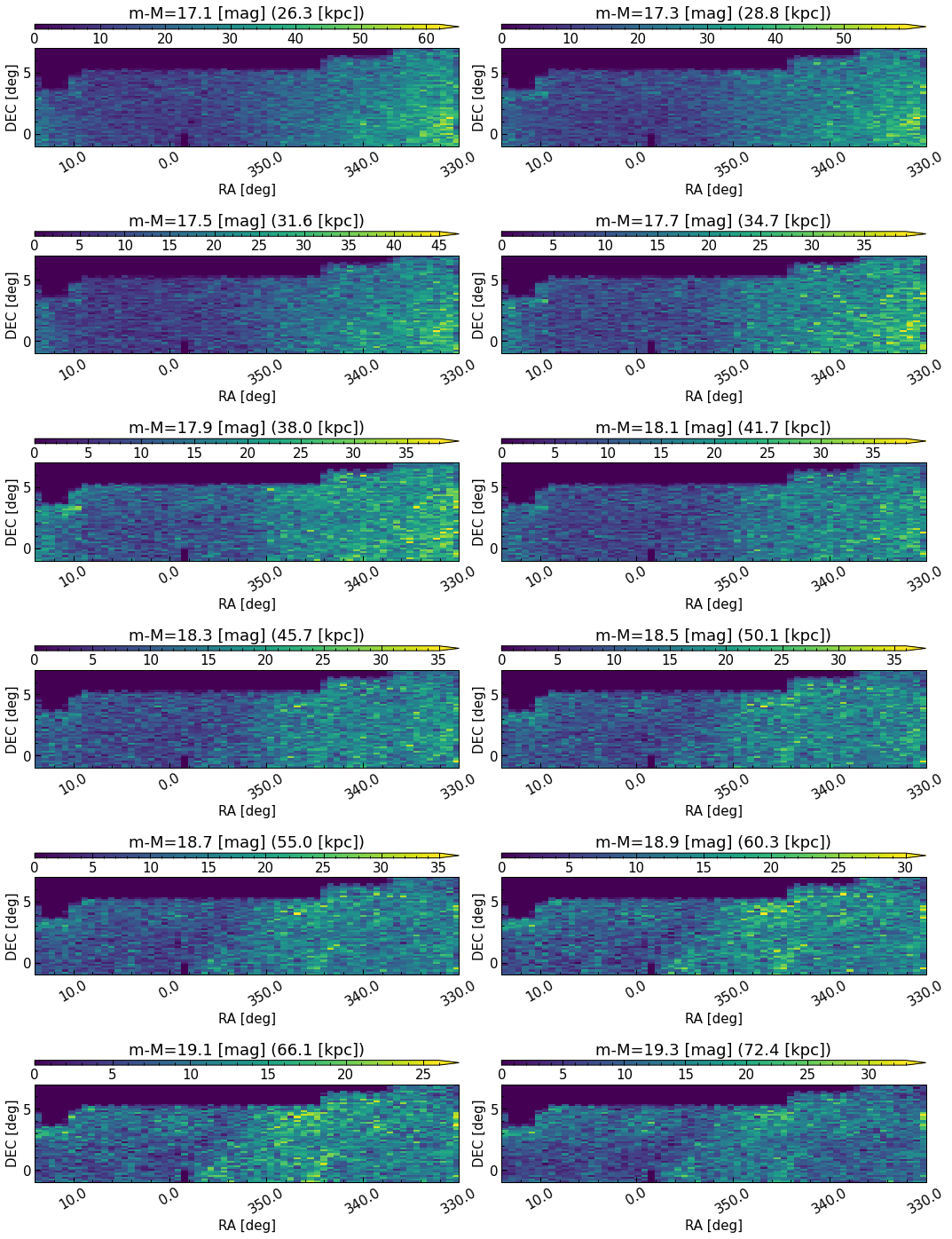}
\end{center}
\hspace*{10mm}
\caption{
Stellar number density distribution in the Fall field after our isochrone-filter selection. This figure shows that there is an excess around RA=330 deg at 30 kpc from the sun. This is the known Helcules-Aquila Cluod. At greater distances, the location of the overdensity region shifts, and a stream-like structure is seen. This corresponds to the Pisces overdensity, and the present study suggests that it may have a stream-like structure and be more spatially extensive than previously thought.
}
\label{Fig10}
\end{figure*}

Here we show how the structure of the Milky Way halo changes with the isochrone-filter method used in this study, i.e., "the tomography of the Milky Way halo". 
We make isochrone filters with distance modulus from 17.1 mag ($D_{\odot}\sim 26.3$ kpc) to 19.3 ($D_{\odot}\sim 72.4$ kpc) mag at the interval of 0.2 mag. The tomography in the Spring, North, and Fall fields are shown in Fig.\ref{Fig8}, \ref{Fig9}, and \ref{Fig10}, respectively.

\section{Effect of the criteria for the star-galaxy separation in the North field}
\begin{figure*}[h!]
\begin{center}
\includegraphics[width=150mm]{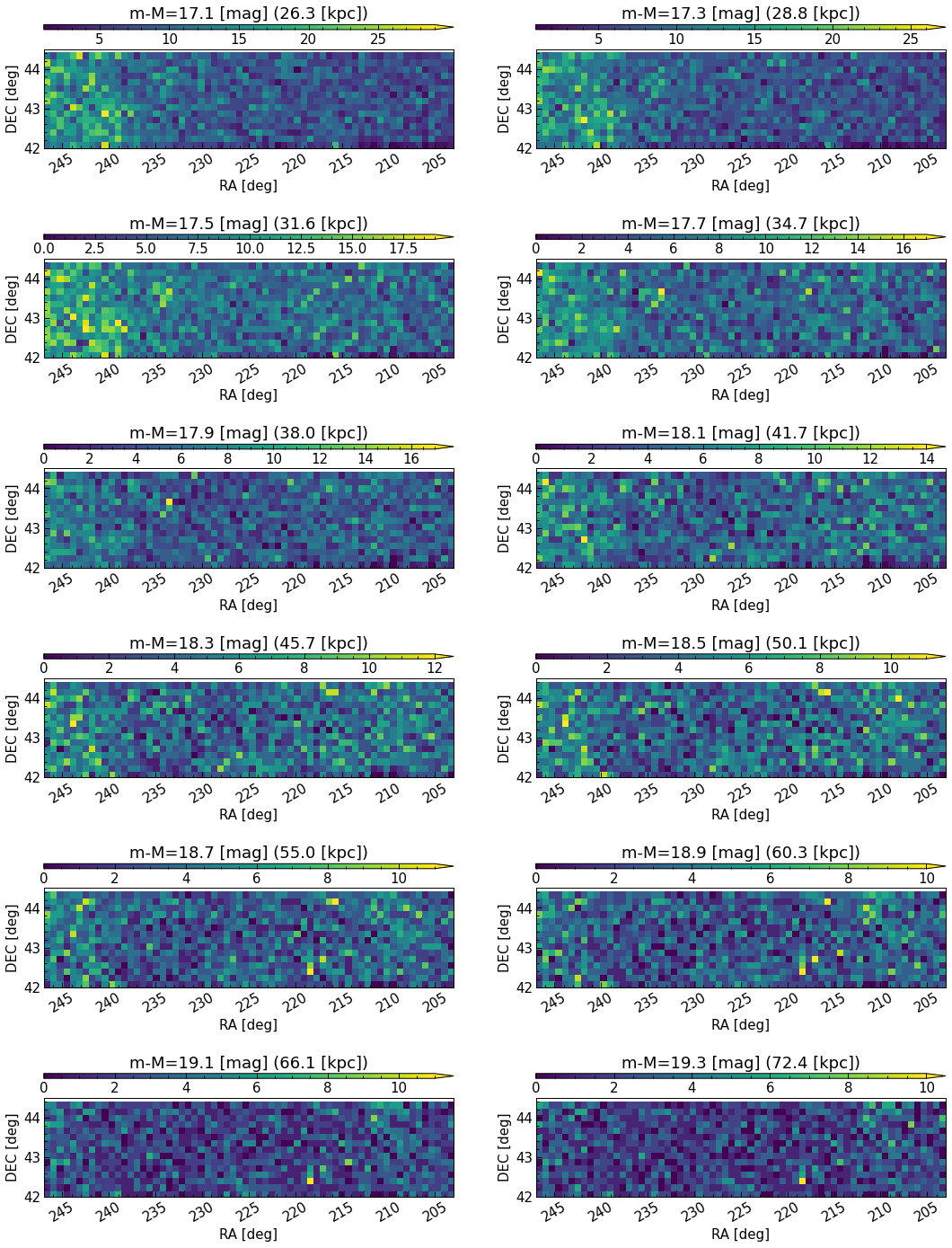}
\end{center}
\hspace*{10mm}
\caption{
Stellar number density distribution in the North field for the stars with $i$ $\leqq$ 24.0 mag. The panels are shown in the same way as Fig.\ref{Fig8}. From this figure, it can be confirmed that the overdensity region around RA=210 deg seen in Fig.\ref{Fig9} disappears by changing the criteria for star-galaxy separation. This suggests that this structure is caused by contamination from distant galaxies. Note, however, that this result does not rule out the existence of B\"ootes overdensity, which is clearly indicated on the color-magnitude diagram.
}
\label{Fig11}
\end{figure*}

We show how the same isochrone-filter results change when only objects brighter than 24 mag in the $i$-band are extracted to investigate the effect of star/galaxy separation. In particular, we show the case for the North field in Fig.\ref{Fig11}, where the effect of this selection is particularly pronounced.

\end{document}